\documentclass[sigconf]{acmart}
\renewcommand\footnotetextcopyrightpermission[1]{} 
\usepackage{booktabs} 
\usepackage{graphicx}
\usepackage{booktabs}
\usepackage{threeparttable}
\usepackage{subfigure}
\usepackage{natbib}
\usepackage{multirow}

\usepackage{enumitem}
\setlist{leftmargin=5.5mm}

\setcopyright{none}
\newcommand{\AceKG}{AceKG}
\newcommand{\AK}{AK}
\acmDOI{10.475/123_4}

\acmISBN{123-4567-24-567/08/06}

\acmConference[CIKM'18]{ACM CIKM conference}{22-26 October 2018}{Lingotto, Turin, Italy}

\begin{document}
\title{\AceKG: A Large-scale Knowledge Graph for Academic Data Mining}

 \author{Ruijie Wang, Yuchen Yan, Jialu Wang, Yuting Jia, Ye Zhang, Weinan Zhang, Xinbing Wang}
 \affiliation{%
   \institution{Shanghai Jiao Tong University, Shanghai, China}
   \postcode{200240}
 }
 \email{{wjerry5, wnzhang, xwang8}@sjtu.edu.cn}

\renewcommand{\shortauthors}{}

\begin{abstract}
Most existing knowledge graphs (KGs) in academic domains suffer from problems of insufficient multi-relational information, name ambiguity and improper data format for large-scale machine processing.
In this paper, we present \AceKG, a new large-scale KG in academic domain. \AceKG~ not only provides clean academic information, but also offers a large-scale benchmark dataset for researchers to conduct challenging data mining projects including link prediction, community detection and scholar classification.
Specifically, \AceKG~describes 3.13 billion triples of academic facts based on a consistent ontology, including necessary properties of papers, authors, fields of study, venues and institutes, as well as the relations among them.
To enrich the proposed knowledge graph, we also perform entity alignment with existing databases and rule-based inference.
Based on \AceKG, we conduct experiments of three typical academic data mining tasks and evaluate several state-of-the-art knowledge embedding and network representation learning approaches on the benchmark datasets built from \AceKG.
Finally, we discuss several promising research directions that benefit from \AceKG.
\end{abstract}


%
%
\maketitle

\section{Introduction}

Knowledge graphs have become very crucial resources to support many AI related applications, such as graph analytics, Q\&A system, web search, etc.
A knowledge graph, which describes and stores facts as triples, is a multi-relational graph consisting of entities as nodes and relations as different types of edges.
Nowadays, many companies and research teams are trying to organize the knowledge in their domain into a machine-readable knowledge graph, e.g., YAGO \cite{YAGO}, NELL \cite{NELL-aaai15}, DBpedia \cite{dbpedia-swj}, and DeepDive \cite{DeepDive}.
Although these large-scale knowledge graphs have collected tremendous amount of factual information about the world, many fields still remain to be covered.

With information of papers, scholars, institutes, venues, fields of study and other useful entities, data mining on academic networks aims to discover hidden relations and to find semantic-based information.
Several academic databases or knowledge graphs have been built with structured academic data \cite{dblp,mag,aminer}. 
The public academic knowledge graphs can provide scholars with convincing academic information, and offer large-scale benchmark datasets for researchers to conduct data mining projects.

However, there are some limitations in existing databases or knowledge graphs. 
First, most of existing works provide homogeneous academic graphs, while relations among different types of entities remaining lost \cite{dblp,aminer}. 
Second, some databases only concentrate on one specific field of study, limiting the projects which aim at discovering cross-field knowledge \cite{dblp}. 
Third, synonymy and ambiguity are also the restrictions for knowledge mining \cite{mag}. 
Allocating the unique IDs to the entities is the necessary solution, but some databases use the names of the entities as their IDs directly.

In this paper, we propose Academic Knowledge Graph (\AceKG), \footnote{http://acemap.sjtu.edu.cn/app/AceKG} 
an academic semantic network, which describes 3.13 billion triples of academic facts based on a consistent ontology, including commonly used properties of papers, authors, fields of study, venues, institutes and relations among them. 
Apart from the knowledge graph itself, we also perform entity alignment with the existing KGs or datasets and some rule-based inferences to further extend it and make it linked with other KGs in the linked open data cloud.
Based on \AceKG, we further evaluate several state-of-the-art knowledge embedding  and network representation learning  approaches in Sections~\ref{sec:ex1} and \ref{sec:ex2}. Finally we discuss several potential research directions that benefit from \AceKG \space in Section \ref{sec:future} and conclude in Section \ref{sec:conclude}.

\begin{figure*}[h]
    \begin{center}
    	\vspace{-8pt}
	\includegraphics[width=0.97\linewidth]{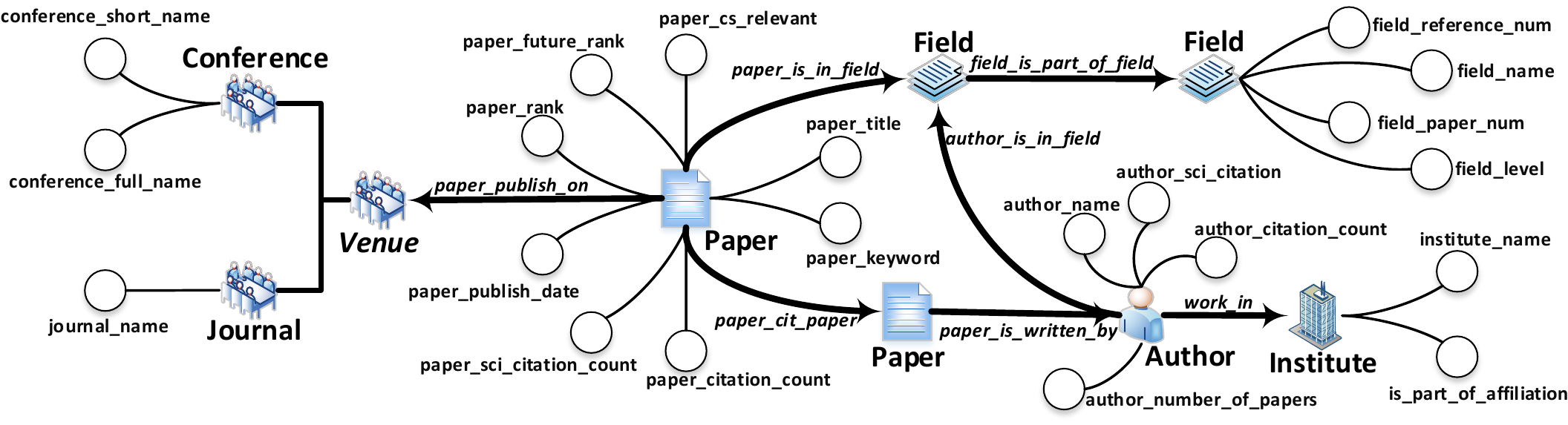}
        \caption{An overview of \AceKG \space Ontology.}
        \label{fig:schema}
        \vspace{-10pt}
    \end{center}
\end{figure*}

Compared with other existing open academic KGs or datasets, \AceKG\space has the following advantages.
\begin{enumerate}
    \item \AceKG\space offers a heterogeneous academic information network, i.e., with multiple entity categories and relationship types, which supports researchers or engineers to conduct various academic data mining experiments.
    \item \AceKG\space is sufficiently large (3.13 billion triples with nearly 100G disk size) to cover most instances in the academic ontology, which makes the experiments based on \AceKG\space more convincing and of practical value.
    \item \AceKG\space provides the entity mapping to computer science databases including ACM, IEEE and DBLP, which helps researchers integrate data from multiple databases together to mine knowledge.
    \item \AceKG\space is fully organized in structured triples, which is machine-readable and easy to process.
\end{enumerate}

\section{The Knowledge Graph} \label{sec:dataset}

The \AceKG\space dataset\footnote{The sample dataset and experiment code are attached in the supplementary material. https://tinyurl.com/CIKM2018-591} can be freely accessed online.
All the data are collected from Acemap\footnote{http://acemap.sjtu.edu.cn}.
\AceKG\space is a large academic knowledge graph with 3.13 billion triples.
It covers almost the whole academic area and offers a heterogeneous academic network.

\subsection{Ontology}
All objects (e.g., papers, institutes, authors) are represented as entities in the \AceKG. Two entities can stand in a relation.
Commonly used attributes of each entities including numbers, dates, strings and other literals are represented as well. 
Similar entities are grouped into classes.
In total, \AceKG\space defines 5 classes of academic entities: \texttt{Papers}, \texttt{Authors}, $\texttt{Fields}\ \texttt{of}\ \texttt{study}$, \texttt{Venues} and \texttt{Institutes}. 
And the facts including the frequently used properties of each entities  and the relations between the entities are described as triples in the knowledge graph.
The ontology of \AceKG\space is shown in Figure \ref{fig:schema}. 

To deal with synonymy and ambiguity, each entity in defined classes are allocated with a URI. 
For example, \texttt{ace:7E7A3A69} and \texttt{ace:7E0D6766} are two scholars having the same name: Jiawei Han, one of whom is the influential data mining scientist. 
Compared with the datasets which use entity names to represent entities directly, \AceKG\space can avoid mistakes caused by synonymy and ambiguity,

The statistics of \AceKG\space are shown in Table \ref{tb:number}. 
All the facts are represented as $subject$-$predicate$-$object$ triples (SPO triples). And we release the Turtle format \AceKG\space online. 
It can be queried by Apache Jena framework\footnote{https://jena.apache.org} with SPARQL easily. 
\begin{table}[t] 
\caption{Triple statistics of \AceKG.}\label{tb:number}
\begin{tabular}{|cr|cr|}\hline
    Class & Number & Class & Number \\\hline
    Paper & 61,704,089& Institute & 19,843 \\
    Author & 52,498,428 & Field & 50,233 \\
    Journal & 21,744 & Conference & 1,278 \\\hline
    Total Entities & 114,295,615 & Total Relations & 3,127,145,831 \\\hline
\end{tabular}
\end{table}

\subsection{Entity alignment}
In order to make \AceKG\space more connected and comprehensive,
we map a large part of papers in computer science of \AceKG\space to the papers stored in IEEE, ACM and DBLP databases.
All the latest papers in those three databases have been aligned with \AceKG\space.
Some mapping statistics are shown in Table \ref{tb:mapping}.
The knowledge graph is updated with the latest academic information periodically. 

\begin{table}[t]\vspace{5pt}
\caption{Statistics of node mapping.}\label{tb:mapping}
\begin{tabular}{|c|ccc|}\hline
    Database & IEEE & ACM & DBLP \\\hline
    Mapping number & 2,332,358 & 1,912,535 & 2,274,773 \\\hline
\end{tabular}\vspace{10pt}
\end{table}

\subsection{Inference}
\begin{figure}[t]
	\centering
	\includegraphics[width=0.4\textwidth]{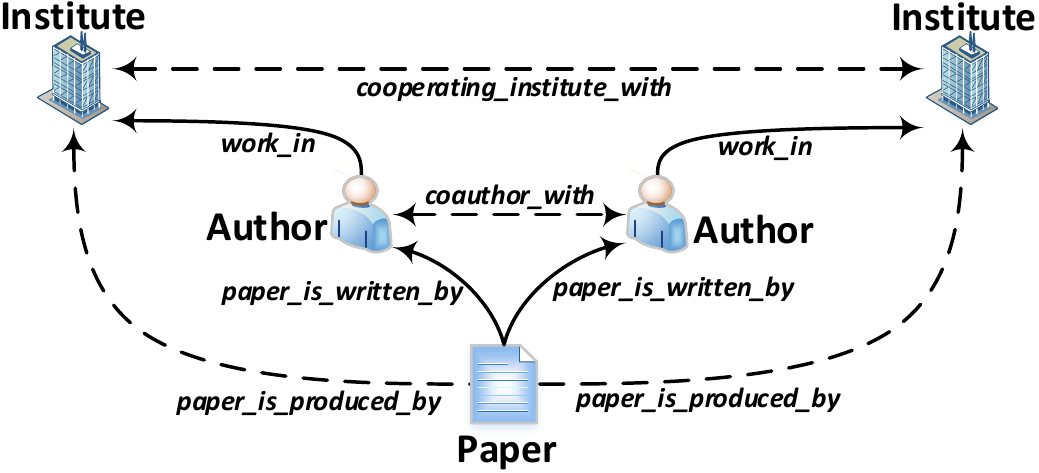}
	\caption{Example of rule-based inference. The dotted arrows are inferred predicates.}
	\label{fig:inference}
	\vspace{10pt}
\end{figure}

Rule-based inference on knowledge graph is a typical but critical way to enrich the knowledge graph.
The selected inference rules that we design are shown in Figure \ref{fig:inference}.
With those inference rules, we can define the new relations on \AceKG, which provides more comprehensive ground truth.

\section{Knowledge embedding} \label{sec:ex1}
In this section, we will evaluate several state-of-the-art approaches for knowledge embedding using \AceKG.
\subsection{Task Definition}
Given a set $S$ of triples $(h, r, t)$ composed of two entities $h, t \in E$ (the set of entities) and a relation $r \in R$ (the set of relationships), knowledge embedding maps each entity to a $k$-dimensional vector in the embedding space, and defines a scoring function to evaluate the plausibility of the triple $(h, r, t)$ in the knowledge graph.
We study and evaluate related methods on link prediction problem defined in \cite{TransE}: given one of the entities and the relation in a latent triple, the task is to predict the other missed entity. 
The commonly used benchmark datasets are FB15K and WN18, which are extracted from Freebase \cite{freebase} and WordNet \cite{wordnet}.
We construct a new benchmark dataset (denoted as \AK18K in the rest of this section) extracted from \AceKG\space for knowledge embedding.
We will show how it differs from FB15K and WN18 in Section \ref{sec:AK18K}.
We compare the following algorithms in our experiments: {TransE} \cite{TransE}, {TransH} \cite{TransH}, {DistMult} \cite{DistMult}, {ComplEx} \cite{ComplEx}, {HolE} \cite{HolE}.

\subsection{Experimental Setup}  \label{sec:AK18K}
To extract \AK18K from \AceKG, we first select 68 critical international venues (conferences and journals) and influential papers published on them.
Then we add the triples of authors, fields and institutes.
Finally, the train/valid/test datasets are divided randomly.

Table \ref{table:kg1} shows the statistics of the WN18, FB15K and \AK18K.
\AK18K is sparser than FB15K but denser than WN18 (indicated by the value of $\#Trip/\#E$), and it provides only 7 types of relations. We will evaluate the models' scalability on the knowledge graph which has simple relation structure but tremendous amount of entities.
The code we used is based on the OpenKE \cite{openke}, an open-source framework for knowledge embedding. 
\begin{table}[t]
\centering
\caption{Datasets used in knowledge embedding.}
\label{table:kg1}
\begin{tabular}{c|ccccc}
\hline
Dataset & \#R & \#E & \multicolumn{3}{c}{\#Trip. (Train/ Valid/ Test)} \\ \hline
WN18 & 18 & 40,943 & 141,442 & 5,000 & 5,000 \\
FB15K & 1,345 & 14,951 & 483,142 & 50,000 & 59,071 \\
\AK18K & 7 & 18,464 & 130,265 & 7,429 & 7,336 \\ \hline
\end{tabular}\vspace{5pt}
\end{table}
\subsection{Evaluation Results}
We show the link prediction results based on knowledge embedding in Table \ref{ke:result}.
The reported results are produced with the best set of hyper-parameters after the grid searches reported in the papers.  
\begin{table}[t]
\begin{center}
\vspace{0pt}
\caption{Results of link prediction task on \AK18K.}
\label{ke:result}
\begin{tabular}{lcccccc} 
\toprule[0.1em]
& \multicolumn{2}{c}{MRR} &  & \multicolumn{3}{c}{Hits at}\\
\cline{2-3} \cline{5-7} Model & Raw & Filter & & 1& 3 & 10 \\
\hline
TransE &0.358&0.719 & & 62.7 &82.5 &\textbf{89.2} \\
TransH & 0.315& 0.701& & 61.0&77.2 &84.6 \\
DistMult & 0.432& 0.749& & 68.7& 79.5&86.1 \\
HolE & \textbf{0.482} &\textbf{0.864} & & \textbf{83.8} & \textbf{87.1}& 88.2\\
ComplEx & 0.440& 0.817& &75.4 &85.8 &89.0 \\
\bottomrule[0.1em]
\end{tabular}
\end{center}
\begin{tablenotes}

\small \item[*] Table note: Filtered and Raw Mean Reciprocal Rank (MRR) and Hits@\{1,3,10\} for the models tested on the AK18K dataset.
Hits@\{1,3,10\} metrics are filtered.
Filtered metrics means removing from the test list the other triples that appear in the dataset while evaluation.
\end{tablenotes}
\vspace{5pt}
\end{table}
The compared state-of-the-art models can be divided into two categories: (i) translational models (TransE, TransH); (ii) compositional models (DistMult, HolE, ComplEx).
TransE outperforms all counterparts on hit@10 as 89.2\%.
Although 94.4\% of relations in our knowledge graph are many-to-many, which works for TransH, TransE shows its advantages on modeling sparse and simple knowledge graph, while TransH fails to achieve better results. The reason may be the number of relationship types is only 7, which is small.
On the other hand, HolE and ComplEx achieve the most significant performances on the other metrics, especially on hit@1 (83.8\% and 75.4\%) and on filtered MRR (0.482 and 0.440), which confirms their advantages on modeling antisymmetric relations because all of our relations are antisymmetric, such as \texttt{field\_is\_part\_of} and \texttt{paper\_is\_written\_by}.

Compared with the experiment results on FB15K and WN18 reported in \cite{HolE}, performances evaluated using \AK18K are noticeably different.
First, results on \AK18K are lower than those on WN18 but higher than those on FB15K. It is caused by the limited relation types and large amount of potential entities per relation. 
Some relation such as \texttt{paper\_is\_in\_field} can have thousands of possible objects per triple, limiting the prediction performance.
Second, the performance gap between two model categories grows more pronounced as the knowledge graph become more complicated, which indicates the translational models with simple assumptions may not model the complicated graph well.

\section{Network representation learning} \label{sec:ex2}
In this section, we will evaluate several state-of-the-art approaches for network representation learning (NRL) on \AceKG.

\subsection{Task Definition}
Given a network $G=(V,E,A)$, where $V$ denotes the vertex set, $E$ denotes the network topology structure and $A$ preserves node attributions, the task of NRL is to learn a mapping function $f : v \mapsto r_v \in R_d$, where $r_v$ is the learned representation of vertex $v$ and $d$ is the dimension of $v_r$.
We study and evaluate related methods including {DeepWalk} \cite{DeepWalk}, {PTE} \cite{PTE}, {LINE} \cite{LINE} and {metapath2vec} \cite{metapath2vec} on two tasks: scholar classification and scholar clustering.

\subsection{Experimental Setup}
Based on \AceKG, we first select 5 fields of study (FOS)\footnote{5 fields of study: Biology, Computer science, Economics, Medicine and Physics.} and 5 main subfields of each.
Then we extract all scholars, papers and venues in those fields of study respectively to construct 5 heterogeneous collaboration networks.
We also construct 2 larger academic knowledge graph:
(i) we integrate 5 networks above into one graph which contains all the information of 5 fields of study;
(ii) we match the eight categories of venues in Google Scholar\footnote{https://scholar.google.com/citations?view\_op=top\_venues\&hl=en\&vq=eng} to those in \AceKG. 151 of 160 venues (8 categories $\times$ 20 per category) are successfully matched. Then we select all the related papers and scholars to construct one large heterogeneous collaboration networks.
The statistics of these networks are shown in Table \ref{tb:dataset_exp23}.
Moreover, the category of scholars are labeled with the following approach:
\begin{enumerate}
    \item To label the papers, we adopt the field of study information and Google scholar category directly as the label of papers in 6 FOS networks and 1 Google scholar network respectively.
    \item As for the label of the scholars, it is determined by the majority of his/her publications' labels. When some labels have equal quantity of papers, they are chosen randomly.
\end{enumerate}

\begin{table}[t]
\centering
\caption{Datasets used in network representation learning.}
\label{tb:dataset_exp23}
\begin{tabular}{c|cccc}
\hline
Dataset & \#Paper & \#Author & \#Venue & \#Edge \\
\hline
FOS\_Biology & 1,211,664 & 2,169,820 & 13,511 & 5,544,376 \\
FOS\_CS & 452,970 & 738,253 & 10,726 & 1,658,917 \\
FOS\_Economics & 412,621 & 597,121 & 8,269 & 1,163,700 \\
FOS\_Medicine & 182,002 & 491,447 & 7,251 & 819,312 \\
FOS\_Physics & 449,844 & 596,117 & 5,465 & 1,602,723 \\
\hline
FOS\_5Fields & 2,578,185 & 3,868,419 & 18,533 & 10,160,137 \\
Google & 600,391 & 635,585 & 151 & 2,373,109 \\
\hline
\end{tabular}
\vspace{5pt}
\end{table}


\subsection{Evaluation Results}
\subsubsection{Classification} \label{sec:cls}

We adopt logistic regression to conduct scholar classification tasks.
Note that in this task 5-fold cross validation are adopted.
Table \ref{tb:cla} shows the classification results evaluated by Micro-F1 and Macro-F1.
metapath2vec learns heterogeneous node embeddings significantly better than other methods.
We attribute it to the modified heterogeneous sampling and skip-gram algorithm.
However, DeepWalk and LINE also achieve comparable performance, showing their scalability on heterogeneous networks.
Another reason for the comparable performance is that our edge types and node types are limited, thus algorithms on homogeneous information network can also learn a comprehensive network representation.

\begin{table*}[htbp]
\centering
\vspace{-0pt}
\caption{Results of scholar classification.}
\label{tb:cla}
\begin{tabular}{c|cccccccc}
\hline
Metric & Method & FOS\_BI & FOS\_CS & FOS\_EC & FOS\_ME & FOS\_PH & FOS\_5F & Google \\ \hline
\multirow{4}{*}{Micro-F1} & DeepWalk & 0.792 & 0.545 & 0.692 & 0.663 & 0.774 & 0.731 & 0.948 \\
 & LINE(1st+2nd) & 0.722 & 0.633 & 0.717 & 0.701 & 0.779 & 0.755 & 0.955 \\
 & PTE & 0.759 & 0.574 & 0.654 & 0.694 & 0.723 & 0.664 & 0.966 \\
 & metapath2vec & 0.828 & 0.678 & 0.753 & 0.770 & 0.794 & 0.831 & 0.971 \\ \hline
\multirow{4}{*}{Macro-F1} & DeepWalk & 0.547 & 0.454 & 0.277 & 0.496 & 0.592 & 0.589 & 0.942 \\
 & LINE(1st+2nd) & 0.445 & 0.542 & 0.385 & 0.577 & 0.640 & 0.655 & 0.949 \\
 & PTE & 0.495 & 0.454 & 0.276 & 0.555 & 0.571 & 0.528 & 0.961 \\
 & metapath2vec & 0.637 & 0.570 & 0.485 & 0.659 & 0.635 & 0.682 & 0.968 \\ \hline

\end{tabular}
\end{table*}

\begin{table}[htbp]
\caption{Results of scholar clustering.}
\label{nrl:cls}
\begin{tabular}{l|cc} 
\toprule[0.1em]
Model & FOS-labeled & Google-labeled \\
\hline
DeepWalk &0.277 &0.394 \\
LINE(1st+2nd) &0.305 & 0.459\\
PTE & 0.153& 0.602\\
metapath2vec &0.427 &0.836 \\
\bottomrule[0.1em]
\end{tabular}
\begin{tablenotes}
\small \item[*]
\end{tablenotes}
\end{table}

It should be noted that there is significant performance gap between FOS-labeled datasets and the Google-labeled dataset, which is because of the different distribution of papers and scholars.
Papers collected in the Google-labeled dataset are published in Top-venues and consequently few scholar could be active in multiple categories, while there are more cross-field papers and scholars in FOS-labeled datasets.

Moreover, the performance indicates the level of interdiscipline in these fields.
For example, the highest Micro-F1 shows that the sub-fields of Biology are the most independent, while the lowest Micro-F1 means that the sub-fields of CS cross mostly.
Finally, the dramatical decline from Micro-F1 to Macro-F1, especially in Economy, indicates the imbalance of sub-fields in some FOS.

\subsubsection{Clustering} 
Based on the same node representation in scholar classification task, we further conduct scholar clustering experiment with k-means algorithm to evaluate the models' performance.
All clustering experiments are conducted 10 times and the average performance is reported.

Table \ref{nrl:cls} shows the clustering results evaluated by normalized mutual information (NMI). Overall, metapath2vec outperforms all the other models, indicating the modified heterogeneous sampling and skip-gram algorithm can preserve the information of the knowledge graph better. 
Another interesting result is the performance gap between FOS-labeled dataset and Google-labeled dataset, which indicates the hypothesis we proposed in section \ref{sec:cls}.

\section{Future directions} \label{sec:future}
There are other research topics which can leverage \AceKG. 
In this section, we propose three potential directions.

\noindent\textbf{Cooperation prediction.}
To predict a researcher's future cooperation behavior is an interesting topic in academic mining, and many current works have contributed to it by considering  previous cooperators, neighborhood, citation relations and other side information.
However, all these factors can be thought as obvious features in an academic knowledge graph, which is incomplete and may always ignore some other features like the same institution or the same field.
Given this situation, one may perform cooperation prediction based on the NRL results, which can represent the features of a researcher better and may provide some help to cooperation prediction task.
 
\noindent\textbf{Author disambiguation.}
Author disambiguation is a traditional problem in social network, which means distinguishing two  people with the same name in a network.
With the help of \AceKG, author disambiguation can be conducted conveniently. 
The network structure and node attributes in \AceKG\space can enhance the author disambiguation performance.
Then, some author disambiguation algorithms with good performance can be applied to  \AceKG.
The author disambiguation problem can be solved and the quality of \AceKG\space will be improved in such an iterative way.  

\noindent\textbf{Finding rising star.}
Finding academic rising star is important in academic mining in that it can provide helpful reference for universities and companies to hire young faculty or new scientist.
Researchers have  raised various algorithms for this based on publication increasing rate, mentoring relations and some other factors.
In order to make the classification better, we can firstly embed the \AceKG\space to uncover the hidden features of rising star and then apply some clustering algorithms on the embedding results.

\section{Conclusion} \label{sec:conclude}
In this paper we propose \AceKG, a large-scale knowledge graph in academic domain, which consists of 3.13 billion triples of academic facts based on a consistent ontology, including commonly used properties of papers, authors, fields of study, venues, institutes and relations among them.
Based on \AceKG, we design three experimental evaluations and further compare several state-of-the-art approaches on \AceKG.
Besides, we propose several potential research topics that can also benefit from the dataset.
We will keep maintaining and updating the coverage of \AceKG\space for wider usage in these directions.

\bibliographystyle{plainnat}
\bibliography{CIKM}

\end{document}